\documentclass[aps,prd,nofootinbib,onecolumn,a4paper]{revtex4-2}
\usepackage{bm,latexsym,amsmath,amssymb,amsfonts,mathrsfs,amsfonts}
\usepackage[dvipdfmx]{graphicx}
\usepackage{comment}
\usepackage{ascmac}
\usepackage{physics}
\usepackage{color,float}
\usepackage{braket}
\usepackage{bbm}
\usepackage{multirow}
\usepackage[dvipsnames]{xcolor}
\newcommand{\simgt}{\lower.5ex\hbox{$\; \buildrel > \over \sim \;$}}
\newcommand{\simlt}{\lower.5ex\hbox{$\; \buildrel < \over \sim \;$}}

\begin{document}
\title{Quantum signature of gravity in optomechanical systems with conditional measurement}
\author{Daisuke Miki,$^{1}$ Akira Matsumura,$^{1}$ Kazuhiro Yamamoto$^{1,2,3}$}
\affiliation{$^1$Department of Physics, Kyushu University, 744 Motooka, Nishi-Ku, Fukuoka 819-0395, Japan}
\affiliation{
$^2$Research Center for Advanced Particle Physics, Kyushu University, 744 Motooka, Nishi-ku, Fukuoka 819-0395, Japan}
\affiliation{
$^3$
International Center for Quantum-field Measurement
Systems for Studies of the Universe and Particles (QUP),
KEK, Oho 1-1, Tsukuba, Ibaraki 305-0801, Japan}
\email{miki.daisuke@phys.kyushu-u.ac.jp, \\
matsumura.akira@phys.kyushu-u.ac.jp, \\
yamamoto@phys.kyushu-u.ac.jp}
\date{\today}

\begin{abstract}
We investigate the quantum signature of gravity in optomechanical systems under quantum control.
We analyze the gravity-induced entanglement and squeezing in mechanical mirrors in a steady state.
The behaviors and the conditions for generating the gravity-induced entanglement and squeezing are identified in the Fourier modes of the mechanical mirrors.
The condition of generating the entanglement between the mirrors found in the present paper is more severe than that of the gravity-induced entanglement between output lights.
The gravity-induced entanglement in optomechanical systems is an important milestone towards verifying the quantum nature of gravity, which should be verified in the future.

\end{abstract}
\maketitle
\section{INTRODUCTION}
Quantum superposition states of spacetime are one of the fundamental assumptions of quantum gravity theory, but it has never been verified at all. 
Feynman considered the result of the quantum superposition of gravity force as a thought experiment
\cite{Feyman}.
The recent proposal of the Bose et al.-Marletto-Vedral (BMV) experiment \cite{Bose,Marletto},
which is a test of gravity-induced entanglement,
can be regarded as a modern feasibility study of the thought experiment by Feynman. 
Quantum entanglement is a quantum nonlocal correlation that cannot occur by classical evolution \cite{Horodecki}.
Hence, the gravity-induced entanglement might be the quantum signature of gravity.
However, there are arguments about what the gravity-induced entanglement proposed in the BMV experiment means for the quantum field theory of gravity \cite{Belenchia18,Carney22,Bose22,Hidaka22,Sugiyama22,Sugiyama23qu}.
In this analysis, the gravity force is described by the variables whose Hilbert space is equivalent to the particles
in the limit of Newtonian gravity.
However, gravity-induced entanglement is the natural
result of quantum gravity, which is derived  from the quantum field theory of linearized gravity. 
Further, under the assumptions of unitarity and Lorentz invariance, the gravity-induced entanglement can be a verification of quantized dynamical degrees of freedom in gravitational field theory as discussed by Carney \cite{Carney22}. 

From an experimental point of view, gravity-induced entanglement has never been observed so far, its experimental verification will be an important milestone for the quantum theory of gravity. 
Towards such experiments associated with Newtonian gravity-induced entanglement, the theoretical studies  have been performed (e.g.,  \cite{Qvarfort,Krisnanda,Nguyen,Miki21,Balushi18,Matsumura20,Miao20,Datta21,miki22,Kaku23,Carney21,Matsumura22}).
Especially, the authors of Refs.~\cite{Miao20,Datta21} discussed the quantum signature of gravity with the use of optomechanics. 
Since the gravitational interaction is very weak, 
the quantum state of a more massive object is advantageous to verify the gravity-induced entanglement.
An optomechanical system consists of a macroscopic mirror and cavity lights, which is promising to generate a massive object in the quantum state \cite{Yanbei,Aspelmeyer,Bowen}.
Quantum control of feedback control \cite{Genes,Rudolph22} and quantum filter \cite{Yanbei08,Yanbei09,Bowen20} is the technique to realize a macroscopic mirror in the quantum state close to the ground state.
Recent studies on the quantum states of macroscopic mirrors with quantum control have demonstrated the feasibility of generating the quantum states of mg-scale objects in the near future \cite{matsumoto20,MY,miki23,Sugiyama23,Shichijo23}.

\begin{figure}[bt]
    \centering
    \includegraphics[width=11.cm]{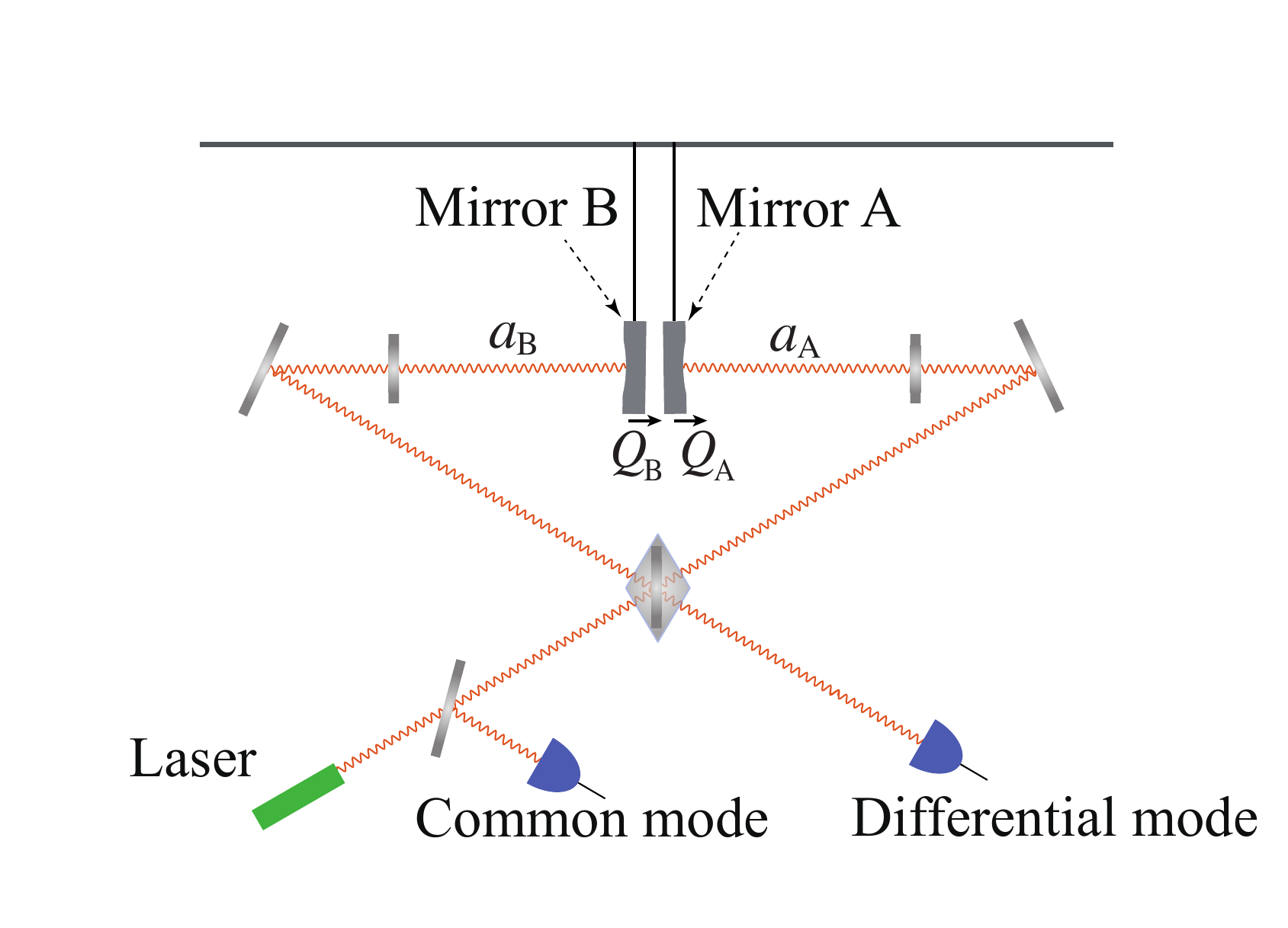}
    \vspace{-3mm}
    \caption{
     A schematic plot of the optomechanical system to generate the gravity-induced entanglement between mirror A and mirror B. We call the differential mode and the common mode of the optical lights as those each coupled to the differential mode and the common mode of the mechanical mode, respectively.}
    \label{fig:GIE}
\end{figure}

In the present paper, we study the quantum signature of Newtonian gravity in optomechanical systems under quantum control.
The authors of Ref.~\cite{Miao20} investigated the gravity-induced entanglement between two output optical modes in the Fourier space, and the authors of Ref.~\cite{Datta21} investigated the gravity-induced squeezing of output lights in the Fourier space.
However, the entanglement and the squeezing of the mechanical modes of mirrors in the optomechanical systems has yet to be investigated.
Therefore, we evaluate the gravity-induced entanglement between mirrors and the gravity-induced squeezing of mirrors under quantum control (See figure \ref{fig:GIE} for a schematic plot of the configuration of the system).
We first derive the condition for generating gravity-induced entanglement between two mirrors without the quantum filter.
We show that this condition is sufficient for generating gravity-induced entanglement between output lights.
Hence, generating gravity-induced entanglement between mirrors found in the present paper is more severe than generating gravity-induced entanglement between lights in Ref.~\cite{Datta21}.
We also demonstrate that this is true even with applying the quantum filter.
We also investigate the behaviors of the gravity-induced squeezing of mechanical modes without the filter.
The squeezing can be useful as a signal of gravity being an operator, but semi-classical gravity also causes similar squeezing \cite{Liu23}.
To this end, we introduce the mechanical quadrature and find the minimum value of the spectral density of the mechanical quadrature in the Fourier space.
We will discuss the behavior of gravity-induced squeezing to examine whether it is useful as a quantum nature of gravity. 

This paper is organized as follows:
In Sec.~II, we briefly review the optomechanical systems under the feedback control, which are gravitationally interacting.
In Sec.~III, we describe the quantum Wiener filter and introduce the mechanical quadratures conditioned on continuous measurements.
In Sec.~IV, we investigate the gravity-induced entanglement between two mirrors in Fourier space.
We also compare the gravity-induced entanglement between two mirrors and that between output light.
In Sec.~V, we focus on the gravity-induced squeezing
of two mirrors. 
Sec.~VI is devoted to summary and conclusions.
In Appendix~A, we present the exact expression for the degree of the entanglement without the quantum filter, which is introduced in Sec.~IV.
Appendix~B explains the optical covariance matrix in Fourier space according to Ref.~\cite{Miao20,Datta21}.
In Appendix~C, we present a brief review of the optical mode squeezing due to gravity found in Ref.~\cite{Datta21}.

\section{FORMULAS}
We consider the cavity optomechanical systems with two mechanical mirrors with mass $m$, as shown in Fig.~\ref{fig:GIE}.
The Hamiltonian of the total system is
\begin{align}
    \label{H}
    \hat{H}&=
    \frac{\hat{P}_A^2}{2m}
    +\frac{1}{2}m\Omega^2\hat{Q}_A^2
    +\frac{\hat{P}_B^2}{2m}
    +\frac{1}{2}m\Omega^2\hat{Q}_B^2
    +\hbar\omega_{c}\hat{a}_{A}^{\dagger}\hat{a}_{A}+\frac{\hbar\omega_{c}}{\ell}\hat{Q}_{A}\hat{a}_{A}^{\dagger}\hat{a}_{A}
    +\hbar\omega_{c}\hat{a}_{B}^{\dagger}\hat{a}_{B}-\frac{\hbar\omega_{c}}{\ell}\hat{Q}_{B}\hat{a}_{B}^{\dagger}\hat{a}_{B}\notag\\
    &\quad
    +i\hbar E\left(e^{-i\omega_{L}t}\frac{\hat{a}_{A}^{\dagger}+\hat{a}_{B}^{\dagger}}{\sqrt{2}}-e^{i\omega_{L}t}\frac{\hat{a}_{A}+\hat{a}_{B}}{\sqrt{2}}\right)+\hat{H}_{g},
\end{align}
where $\hat{Q}_{j}$ and $\hat{P}_{j}$ with $j=A,~B$ are the canonical operators satisfying $[\hat{Q}_{j},\hat{P}_{j}]=i\hbar$,
$\hat{a}_{j}$ and $\hat{a}_{j}^{\dagger}$ with $j=A,~B$ are the annihilation and creation operators of optical cavity modes satisfying $[\hat{a}_{j},\hat{a}_{j}^\dagger]=1$, 
$\omega_{c}$ is the resonance frequency of the cavity lights, and $\ell$ is the cavity length.
The first term in the second line of the Hamiltonian expresses the input laser from the common side, in which 
$\omega_{L}$ is the laser frequency and $E=\sqrt{P_{\text{in}}\kappa/\hbar\omega_{L}}$ is the laser amplitude with the laser power $P_{\text{in}}$ and the optical decay rate of the cavity lights $\kappa$. The last term of the Hamiltonian
$\hat{H}_{g}$ describes the gravitational interaction between the two mirrors given by
\begin{align}
    \label{Hgr}
    \hat{H}_{g}&=
    -\frac{Gm^{2}}{L}\frac{1}{1-(\hat{Q}_{B}-\hat{Q}_{A})/L},
\end{align}
where $G$ is the gravitational constant and $L$ is the distance between two mirrors when they are in equilibrium.

We consider the perturbations around a steady state as $\hat{Q}_j=\bar{Q}_j+\delta\hat{Q}_j,~\hat{P}_j=\bar{P}_j+\delta\hat{P}_j,~\hat{a}_j=\bar{a}_j+\delta\hat{a}_j$, where 
$(\bar{Q}_j,\bar{P}_j,\bar{a}_j)$ are the classical mean values of $(\hat{Q}_j,\hat{P}_j,\hat{a}_j)$ and $(\delta\hat{Q}_j,\delta\hat{P}_j,\delta\hat{a}_j)$ are the perturbative quantities, and then the Hamiltonian is expanded up to the second order of the perturbations.
We introduce the common mode and the differential mode, 
which are denoted by the subscript $+$ and $-$, respectively, as
\begin{align}
    \hat{Q}_\pm&=
    \frac{\hat{Q}_A\pm\hat{Q}_B}{\sqrt{2}},\quad
    \hat{P}_\pm=
    \frac{\hat{P}_A\pm\hat{P}_B}{\sqrt{2}},\quad
    \hat{a}_\pm=
    \frac{\hat{a}_A\mp\hat{a}_B}{\sqrt{2}}.
\end{align}
The optical common (differential) mode is defined as the mode of light coupled with the mechanical common (differential) mode.
We assume the symmetrical setup, which leads to $(\bar{Q}_2,\bar{P}_2,\bar{a}_2)=(-\bar{Q}_1,-\bar{P}_1,\bar{a}_1)$ and $\bar{Q}_+=\bar{P}_+=\bar{a}_+=0$.
The equations for the classical mean values are
\begin{align}
    \label{qbar}
    &\dot{\bar{Q}}_-
    =\frac{\bar{P}_-}{m},\\
    &\dot{\bar{P}}_-
    =-m\Omega^2\bar{Q}_-
    -\frac{\sqrt{2}Gm^2}{L^2}
    -\frac{\hbar\omega_c}{\sqrt{2}\ell}|\bar{a}_-|^2
    -\Gamma\bar{P}_-,\\
    &\dot{\bar{a}}_-
    =i\Delta\bar{a}_-+E-\frac{\kappa}{2}\bar{a}_-,
    \label{abar}
\end{align}
where we replaced $\hat{a}$ by $\hat{a}e^{i\omega_Lt}$ and introduced the detuning $\Delta=\omega_{L}-\omega_{c}(1+\bar{Q}_-/\ell)$ and the mechanical dissipation rate $\Gamma$.
For the perturbations, instead of $\delta \hat{a}_\pm$ and $\delta \hat{a}^\dagger_\pm$,  we adopt the optical amplitude quadratures $\delta\hat{x}_\pm=e^{-i\theta}\delta\hat{a}_\pm+e^{i\theta}\delta\hat{a}_\pm^{\dagger}$ and the optical phase quadratures $\delta\hat{y}_\pm=(e^{-i\theta}\delta\hat{a}_\pm-e^{i\theta}\delta\hat{a}_\pm^{\dagger})/i$.
The Langevin equations for the perturbative quantities are
\begin{align}
    \dot{\delta\hat{Q}}_{\pm}&=
    \frac{\delta\hat{P}_{\pm}}{m},\\
    \dot{\delta\hat{P}}_{\pm}&=
    -m\Omega_\pm^2\delta\hat{Q}_{+}
    -\frac{\hbar\omega_c|\bar{a}_-|}{\sqrt{2}\ell}\delta\hat{x}_{\pm}
    -\Gamma\delta\hat{P}_{\pm}+\sqrt{2\Gamma}\hat{P}_{\pm}^{\text{in}}
    -\int_{-\infty}^{t}dsg_{\text{FB}}(t-s)\hat{I}(s)
    \label{feedback}
    ,\\
    \dot{\delta\hat{x}}_{\pm}&=
    -\frac{\kappa}{2}\delta\hat{x}_{\pm}-\Delta\delta\hat{y}_{\pm}+\sqrt{\kappa}x_{\pm}^{\text{in}},\\
    \dot{\delta\hat{y}}_{\pm}&=
    -\frac{\sqrt{2}\omega_c}{\ell}\delta\hat{Q}_{\pm}
    +\Delta\delta\hat{x}_{\pm}-\frac{\kappa}{2}\delta\hat{y}_{\pm}+\sqrt{\kappa}y_{\pm}^{\text{in}},\end{align}
where we defined
\begin{align}
    &\Omega_+=\Omega,\quad
    \Omega_-=\Omega\sqrt{1-\epsilon},\\
    \epsilon
    &=\frac{4Gm}{L^{3}\Omega^{2}}.
\end{align}
Here we introduced the noise terms
$\hat{P}_\pm^{\text{in}}$,  $\hat{x}_\pm^{\text{in}}$, and $\hat{y}_\pm^{\text{in}}$, as well as the feedback force in Eq.~\eqref{feedback}, where $g_\text{FB}$ describes the feedback control and $\hat{I}$ is the optical output quadrature \cite{Genes,MY}.
The noises have the zero mean values,  $\braket{\hat{P}_\pm^{\text{in}}}=0$, 
$\braket{\hat{x}_\pm^{\text{in}}}=0$, and 
$\braket{\hat{y}_\pm^{\text{in}}}=0$, 
and must satisfy the dissipation-fluctuation relation.
For the phonon noises of the mirrors, we assume
\begin{align}
    \braket{\hat{P}_\pm^{\text{in}}(t)\hat{P}_\pm^{\text{in}}(t')}&=
    \int\frac{d\omega}{2\pi}\frac{m\hbar\omega}{2}
    e^{-i\omega(t-t')}
    \left[\coth(\frac{\hbar\omega}{2k_BT})-1\right],
\end{align}
where $k_B$ is the Boltzmann's constant and $T$ is the environmental temperature. For the optical noise input,
$\hat{x}_\pm^{\text{in}}$ and $\hat{y}_\pm^{\text{in}}$,
we assume
\begin{align}
    \braket{\hat{x}_\pm^{\text{in}}(t)\hat{x}_\pm^{\text{in}}(t')}&=
    \braket{\hat{y}_\pm^{\text{in}}(t)\hat{y}_\pm^{\text{in}}(t')}=
    (2N_{\text{th}}+1)\delta(t-t'),
\end{align}
where $N_{\text{th}}=(e^{\hbar\omega_c/k_BT}-1)^{-1}$ is the thermal photon number, which can be regarded as $N_{\text{th}}\simeq0$ because of the high optical frequency $\hbar\omega_c/k_BT\gg1$.

Introducing the dimensionless operators as
\begin{align}
    \hat{q}_\pm&=
    \sqrt{\frac{2m\Omega_\pm}{\hbar}}\hat{Q}_\pm,\quad
    \hat{p}_\pm=
    \sqrt{\frac{2}{m\hbar\Omega_\pm}}\hat{P}_\pm,
\end{align}
which satisfy $[\hat{q}_\pm,\hat{p}_\pm]=2i$,
the Langevin equations are rewritten as
\begin{align}
    \dot{\delta\hat{q}}_{\pm}&=
    \Omega_\pm\delta\hat{p}_{\pm},\\
    \dot{\delta\hat{p}}_{\pm}&=
    -\Omega_\pm\delta\hat{q}_{+}
    -2g_\pm\delta\hat{x}_{\pm}
    -\gamma_m\delta\hat{p}_{\pm}+\sqrt{2\gamma_m}\hat{p}_{\pm}^{\text{in}},\\
    \dot{\delta\hat{x}}_{\pm}&=
    -\frac{\kappa}{2}\delta\hat{x}_{\pm}
    -\Delta\delta\hat{y}_\pm
    +\sqrt{\kappa}x_{\pm}^{\text{in}},\\
    \dot{\delta\hat{y}}_{\pm}&=
    -2g_\pm\delta\hat{q}_{\pm}
    +\Delta\delta\hat{x}_\pm
    -\frac{\kappa}{2}\delta\hat{y}_{\pm}+\sqrt{\kappa}y_{\pm}^{\text{in}},
\end{align}
where $\gamma_m$ is the mechanical decay rate under feedback control, 
$g_\pm$ is the optomechanical coupling given by
\begin{align}
    g_{\pm}&=
    \frac{\omega_c}{\ell}\sqrt{\frac{\hbar}{2m\Omega_\pm}}\frac{|\bar{a}_-|}{\sqrt{2}}, 
\end{align}
and $\hat{p}_\pm^{\text{in}}$ is the thermal fluctuation whose 
correlation is written as 
\begin{align}
    \frac{1}{2}\braket{\{\hat{p}_\pm^{\text{in}}(t),\hat{p}_\pm^{\text{in}}(t')\}}&=
    \int\frac{d\omega}{2\pi}\frac{\omega}{\Omega_\pm}
    e^{-i\omega(t-t')}
    \coth(\frac{\hbar\omega\gamma_m}{2k_BT\Gamma})
    \simeq
    (2n_{\text{th}}^{\pm}+1)\delta(t-t').
\end{align}
Here $\{\hat{\cal A},\hat{\cal B}\}=\hat{\cal A}\hat{\cal B}+\hat{\cal B}\hat{\cal A}$ denote the anti-commutator,
and $n_{\text{th}}^{\pm}=(e^{\hbar\Omega_\pm\gamma_m/k_BT\Gamma}-1)^{-1}$ is the thermal phonon number.

In the present paper, we consider the measurement of the optical phase quadrature $\delta\hat{y}_{\pm}$.
The input-output relation is given by
\begin{align}
    \hat{y}_\pm^{\text{out}}&=
    \hat{y}_\pm^{\text{in}}-\sqrt{\kappa}\delta\hat{y}_\pm.
\end{align}
Considering the additional vacuum noise input because of the imperfect detection \cite{Genes}, 
the optical output quadrature is
\begin{align}
    \hat{Y}_\pm&=\sqrt{\eta}\hat{y}_\pm^{\text{out}}+\sqrt{1-\eta}\hat{y}_\pm^{\text{in}\prime},
\end{align}
where $\braket{\{\hat{y}_\pm^{\text{in}\prime}(t),\hat{y}_\pm^{\text{in}\prime}(t')\}}=2\delta(t-t')$ and $\eta\in[0,1]$ is the detection efficiency.
We henceforth assume $\Delta=0$, which can be adjusted by slightly varying the cavity length.
We also assume that the mirror is in a steady state, in which case the background classical solution is obtained by $\dot{\bar{Q}}_-=\dot{\bar{P}}_-=\dot{\bar{a}}_-=0$.

\section{Wiener filter for optical phase measurement}
In this section, we consider the Langevin equations in the Fourier domain and perform the quantum Wiener filter analysis to optimally estimate the mechanical motion of the mirrors.
Defining the variabes in Fourier domain as $f (\omega)=\int_{-\infty}^{\infty}dtf(t)e^{i\omega t}$,
we have the solution of the Langevin equations, 
\begin{align}
    \label{delq}
    \delta\hat{q}_\pm(\omega)&=
    \frac{\Omega_\pm}{\Omega_\pm^2-i\gamma_m\omega-\omega^2}
    \left(\sqrt{2\gamma_m}\hat{p}_{\pm}^{\text{in}}(\omega)
    -\frac{2ig_\pm\sqrt{\kappa}}{\omega+i\kappa/2}\hat{x}_\pm^{\text{in}}(\omega)\right),\\
    \delta\hat{p}_\pm(\omega)&=
    -\frac{i\omega}{\Omega_\pm}\delta\hat{q}_\pm(\omega),\\
    \delta\hat{x}_\pm(\omega)&=
    \frac{i\sqrt{\kappa}}{\omega+i\kappa/2}\hat{x}_\pm^{\text{in}}(\omega),\\
    \delta\hat{y}_\pm(\omega)&=
    \frac{i}{\omega+i\kappa/2}
    \left(-2g_\pm\delta\hat{q}_\pm(\omega)+\sqrt{\kappa}\hat{y}_\pm^{\text{in}}(\omega)\right).
    \label{dely}
\end{align}
We obtain the mechanical spectral density as
\begin{align}
    \label{sqq}
    S_{qq}^\pm(\omega)&=
    \frac{\Omega_\pm^2}{(\Omega_\pm^2-\omega^2)^2+\gamma_m^2\omega^2}
\left(2\gamma_m(2n^\pm_{\text{th}}+1)+\frac{4g_\pm^2\kappa}{\omega^2+\kappa^2/4}(2N_{\text{th}}+1)\right),\\
    S_{qp}^\pm(\omega)&=\frac{i\omega}{\Omega_\pm}S_{qq}^\pm(\omega),\quad
    S_{pp}^\pm(\omega)=\frac{\omega^2}{\Omega_\pm^2}S_{qq}^\pm(\omega),
\end{align}
where the spectral density of the operator $\hat{A}$ and $\hat{B}$ are defined as
\begin{align}
    2\pi\delta(\omega-\omega')S_{AB}(\omega)&=
    \frac{1}{2}\braket{\{\hat{A}(\omega),\hat{B}^\dagger(\omega')\}},
\end{align}
and we used the correlation functions of the fluctuation given by
\begin{align}
    \braket{\{\hat{p}_{\pm}^{\text{in}}(\omega),\hat{p}_{\pm}^{\text{in}}(\omega')\}}&=
    2(2n_{\text{th}}^{\pm}+1)\times2\pi\delta(\omega+\omega'),\\
    \braket{\{\hat{x}_{\pm}^{\text{in}}(\omega),\hat{x}_{\pm}^{\text{in}}(\omega')\}}&=
    \braket{\{\hat{y}_{\pm}^{\text{in}}(\omega),\hat{y}_{\pm}^{\text{in}}(\omega')\}}=
    2(2N_{\text{th}}+1)\times2\pi\delta(\omega+\omega').
\end{align}

The optical output quadrature, which has information on the mechanical motion of mirrors, is
\begin{align}
    \hat{Y}_\pm(\omega)&=
    \frac{2ig_\pm\sqrt{\kappa\eta}}{\omega+i\kappa/2}\delta\hat{q}_\pm
    +\frac{\omega-i\kappa/2}{\omega+i\kappa/2}\sqrt{\eta}\hat{y}_\pm^{\text{in}}
    +\sqrt{1-\eta}\hat{y}_\pm^{\text{in}\prime}\\
    &=
    \frac{1}{F_\pm(\omega)}
    \biggl(2ig_\pm\Omega_\pm\sqrt{2\gamma_m\kappa\eta}(\omega+\frac{i\kappa}{2})\hat{p}_\pm^{\text{in}}(\omega)
    +4g_\pm^2\Omega_\pm\kappa\sqrt{\eta}\hat{x}_\pm^{\text{in}}(\omega)\notag\\
    &\qquad
    +F_\pm(\omega)\frac{\omega-i\kappa/2}{\omega+i\kappa/2}\sqrt{\eta}\hat{y}_\pm^{\text{in}}(\omega)+F_\pm(\omega)\sqrt{1-\eta}\hat{y}_\pm^{\text{in}\prime}(\omega)\biggr),
\end{align}
where $F_\pm(\omega)$ is given by
\begin{align}
    F_\pm(\omega)&=
    (\Omega_\pm^2-i\Gamma\omega-\omega^2)\left(\omega+\frac{i\kappa}{2}\right)^2.
\end{align}
We note that 
$F_\pm(\omega)$ is a 
function where the imaginary part of the solution for $F_\pm(\omega)=0$ is negative.
The spectral densities $S_{YY}$, $S_{qY}$, and $S_{pY}$ are computed as
\begin{align}
    S_{YY}^\pm(\omega)&=
    \frac{2\eta N_{\text{th}}+1}{|F_\pm(\omega)|^2}
    \left(
    |F_\pm(\omega)|^2
    +8g_\pm^2\Omega_\pm^2\gamma_m\kappa\eta(\omega^2+\frac{\kappa^2}{4})\frac{2n_{\text{th}}^\pm+1}{2\eta N_{\text{th}}+1}
    +16g_\pm^4\Omega_\pm^2\kappa^2\eta\frac{2N_{\text{th}}+1}{2\eta N_{\text{th}}+1}
    \right),\\
    S_{qY}^\pm(\omega)&=
    \frac{2N_{\text{th}}+1}{|F_\pm(\omega)|^2}
    \left(
    -4ig_\pm\Omega_\pm^2\gamma_m\sqrt{\kappa\eta}(\omega^2+\frac{\kappa^2}{4})(\omega+\frac{i\kappa}{2})\frac{2n_{\text{th}}^\pm+1}{2N_{\text{th}}+1}
    -8ig_\pm^3\Omega_\pm^2\kappa\sqrt{\kappa\eta}(\omega+\frac{i\kappa}{2})
    \right),\\
    S_{pY}^\pm(\omega)&=\frac{-i\omega}{\Omega_\pm}S_{qY}^\pm(\omega).
\end{align}
Introducing the quantum Wiener filters \cite{MY,Bowen20,Shichijo23} as
\begin{align}
    H_q^\pm(\omega)&=
    \frac{1}{[S_{YY}^\pm(\omega)]_{\text{c}}}
    \left[\frac{S_{qY}^\pm(\omega)}{[S_{YY}^\pm(\omega)]_{\text{nc}}}\right]_c,\\
    H_p^\pm(\omega)&=
    \frac{1}{[S_{YY}^\pm(\omega)]_{\text{c}}}
    \left[\frac{S_{pY}^\pm(\omega)}{[S_{YY}^\pm(\omega)]_{\text{nc}}}\right]_c,
\end{align}
where 
$[f(\omega)]_{\text{c}}$ and $[f(\omega)]_{\text{nc}}$ describe the causal and non-causal part of $f(\omega)$, respectively, we have the mechanical quadratures conditioned on the measurement result,
\begin{align}
    \tilde{q}_\pm(\omega)&=\delta\hat{q}_\pm(\omega)-H_q^\pm(\omega)\hat{Y}_\pm(\omega),\\
    \tilde{p}_\pm(\omega)&=\delta\hat{p}_\pm(\omega)-H_p^\pm(\omega)\hat{Y}_\pm(\omega).
\end{align}
The quantum filters 
$H^\pm_q(\omega)$ and $H^\pm_p(\omega)$ minimize the all components of the mechanical covariance matrix \cite{MY,Bowen20,Shichijo23}.
The causal part and the non-causal part of the spectral density $S_{YY}$ are
\begin{align}
    [S_{YY}^\pm(\omega)]_{\text{c}}&=
    \sqrt{2\eta N_{\text{th}}+1}\frac{F_\pm'(\omega)}{F_\pm(\omega)},\quad
    [S_{YY}^\pm(\omega)]_{\text{nc}}=
    \sqrt{2\eta N_{\text{th}}+1}\frac{F_\pm^{\prime*}(\omega)}{F_\pm^*(\omega)},
\end{align}
where $F'_\pm (\omega)$ is the causal function given as the solution of the following equation:
\begin{align}
    |F'_\pm(\omega)|^2&=
    |F_\pm(\omega)|^2
    +8g_\pm^2\Omega_\pm^2\gamma_m\kappa\eta(\omega^2+\frac{\kappa^2}{4})\frac{2n_{\text{th}}^\pm+1}{2\eta N_{\text{th}}+1}
    +16g_\pm^4\Omega_\pm^2\kappa^2\eta\frac{2N_{\text{th}}+1}{2\eta N_{\text{th}}^\pm+1}.
\end{align}

\section{Entanglement in Fourier domain}
In this section, we focus on the gravity-induced entanglement between the individual mirrors.
Ref.~\cite{Mancini} showed the entanglement criterion in the Fourier domain as
\begin{align}
    \label{ende}
    E(\omega)&=
    \frac{\braket{\hat{R}_{\tilde{q}_+}(\omega)^2}\braket{\hat{R}_{\tilde{p}_-}(\omega)^2}\Omega_-/\Omega}{|\braket{[\hat{R}_{\tilde{q}_A}(\omega),\hat{R}_{\tilde{p}_A}(\omega)]}|^2}<1,
\end{align}
where $\hat R_Z(\omega)$ is defined by $\hat R_Z(\omega)=(\hat{Z}(\omega)+\hat{Z}(-\omega))/2$ with $Z=\tilde q_+, ~\tilde p_-,~ \tilde q_A$, and $\tilde p_A$, and 
$\hat R_Z(\omega)$ is the Hermitian operator in Fourier domain.
$\tilde{q}_A(\omega)=(\tilde{q}_+(\omega)+\tilde{q}_-(\omega)\sqrt{\Omega/\Omega_-})/2$ and
$\tilde{p}_A(\omega)=(\tilde{p}_+(\omega)+\tilde{p}_-(\omega)\sqrt{\Omega_-/\Omega})/2$ are the conditional canonical operators of the mechanical mirror A.
According to the reference~\cite{Giovannetti01}, 
the commutation relations of the fluctuations are
\begin{align}
    [\hat{p}_{\pm}^{\text{in}}(\omega),\hat{p}_{\pm}^{\text{in}}(\omega')]&=
    \frac{2\omega}{\Omega_\pm}\times2\pi\delta(\omega+\omega'),\\
    [\hat{x}_{\pm}^{\text{in}}(\omega),\hat{y}_{\pm}^{\text{in}}(\omega')]&=
    2i\times2\pi\delta(\omega+\omega').
\end{align}
Here we note that the commutation relation of $\hat{p}^{\text{in}}(\omega)$ was found in Ref.~\cite{Giovannetti01}.
We also note that the inequality \eqref{ende} is the sufficient condition for the entanglement.

We here define quality factor $Q_\pm$ and cooperativity $C_\pm$ for the optomechanical systems as follows:
\begin{align}
    Q_\pm&=
    \frac{\Omega_\pm}{\gamma_m},\quad
    C_\pm=
    \frac{4g_\pm^2}{\gamma_m\kappa},
\end{align}
where the quality factor describes the dissipation rate of the system, while the cooperativity characterizes the strength of the measurement, and $\gamma_mC_\pm$ represents the measurement rate.
Hereafter, we fix the Fourier frequency to the resonant frequency of the mirror $\omega=\Omega$.
We also assume $\eta=1$ and $N_{\text{th}}=0$ for simplicity.
Then, the entanglement between the mechanical mirrors or between the output lights is determined only by the five dimensionless parameters $(C_+,Q_+,n_{\text{th}}^+,\kappa/\gamma_m,\epsilon)$.
We first consider the case without quantum filtering.
Appendix A shows the analytical expression for the degree of quantum entanglement $E(\Omega)$.
We here consider the entanglement criterion under the two assumptions.
One is that the quantum cooperativity is larger than unity $C_q^+\equiv C_+/n_{\text{th}}^+\gg1$,
which means that the second term of the right-hand side of the spectral density \eqref{sqq} is dominant compared to the first term.
Because the second term is adjusted by controlling lights, this condition guarantees that we can control the motion of the mirrors through the interaction with lights.
Another assumption is the condition of a bad cavity  $\kappa\gg\Omega$, which guarantees that we continuously get information on the mirror's position.
These assumptions are necessary for quantum control and continuous measurements. 
Then we have found that 
the entanglement criterion \eqref{ende} yields
\begin{align}
    Q_+\epsilon>4C_+,
    \label{encr}
\end{align}
which is equivalent to
\begin{align}
    \frac{Gm}{L^3\Omega}>4\frac{g_+^2}{\kappa}.
\end{align}
This inequality represents that the gravitational interaction between the mechanical mirrors is dominant against
the measurement process using the cavity lights. 
This guarantees the generation of the gravity-induced entanglement between the mirrors.

\begin{figure}[tbp]
    \centering
    \includegraphics[width=8.5cm]{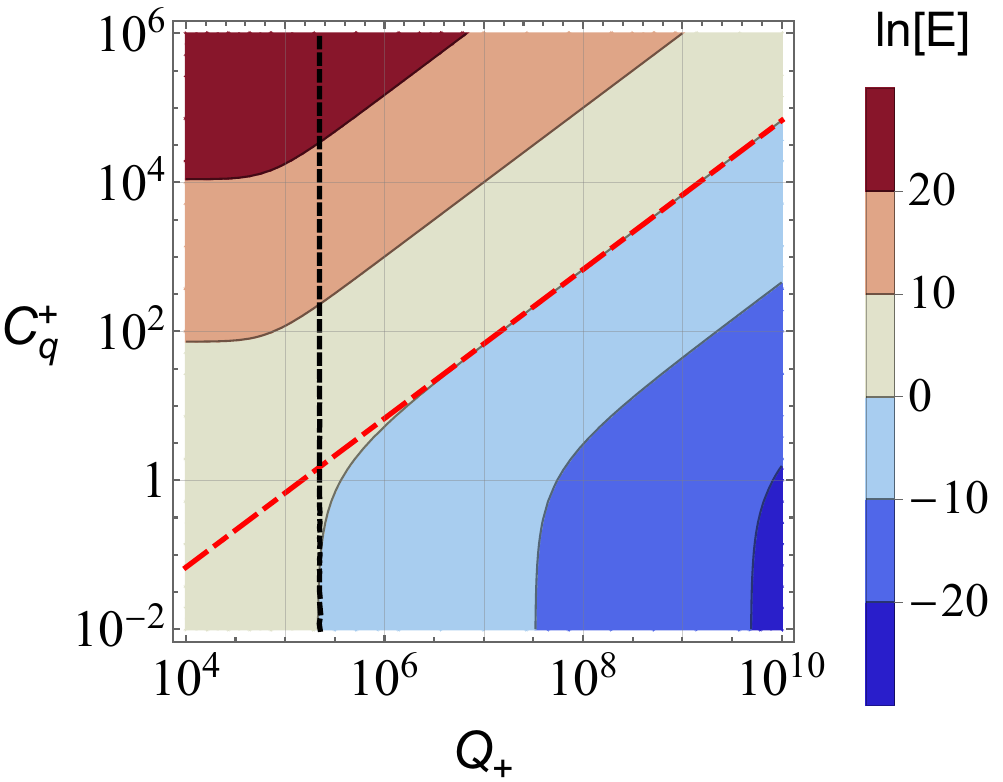}
    \hspace{0.7cm}
    \includegraphics[width=8.5cm]{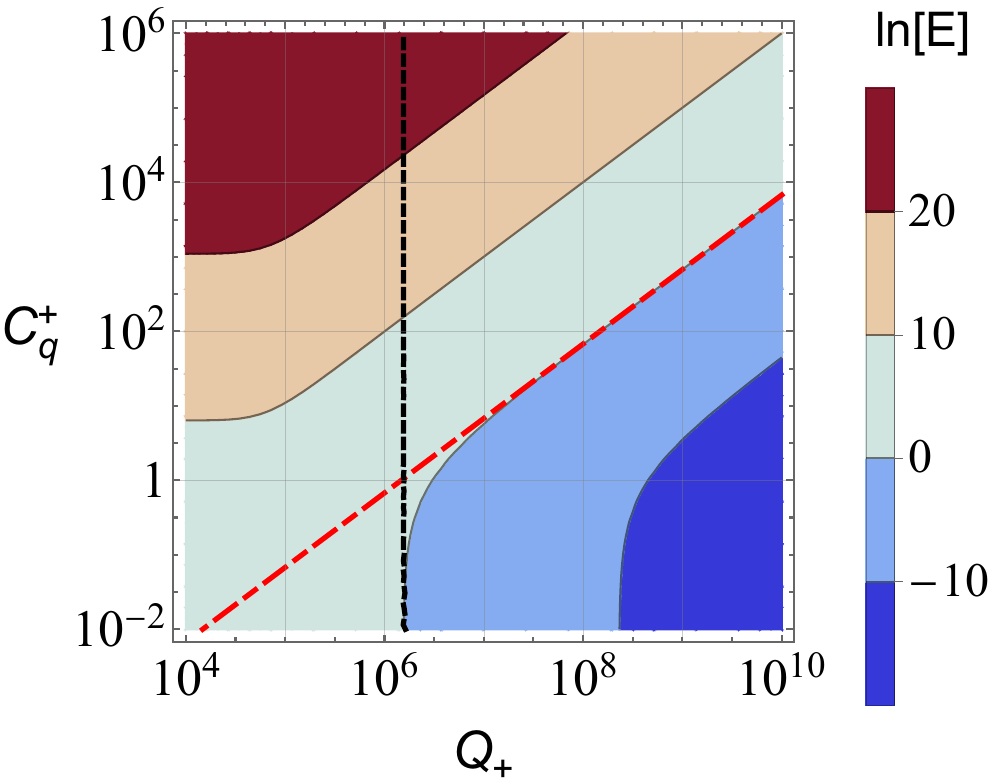}
    \caption{
    Entanglement behavior of $\ln[E(\Omega)]$ as a function of quality factor $Q_+$ and quantum cooperativity $C_q^+=C_+/n_{\text{th}}^+$ without the quantum Wiener filter.
    If $\text{ln}[E(\Omega)]<0$, the two mechanical mirrors are entangled.
    Parameters are
    $\kappa/\gamma_m=10^{13}$,
    $\epsilon=2.7\times10^{-5}$,
    $n_{\text{th}}^+=1$ (left panel),
    and $n_{\text{th}}^+=10$ (right panel).
    The red dashed line represents the boundary of the approximate entanglement condition Eq.~\eqref{encr}.
    Gravity-induced entanglement between output lights occurs in the region to the right of the black vertical dashed line.}
    \label{fig:ennf}
\end{figure}

On the other hand, Ref.~\cite{Miao20,Datta21} showed that the condition for generating the gravity-induced entanglement between output lights is
\begin{align}
    \label{enphcr}
    Q_+\epsilon>4n_{\text{th}}^+,
\end{align}
under the same assumptions.
This is the necessary condition for the inequality \eqref{encr}
as long as $C_q^+\gg1$.
It is therefore more severe to verify the gravity-induced entanglement between the mirrors than that between the output lights, which is clearly demonstrated in Figure \ref{fig:ennf}.

Fig.~\ref{fig:ennf} shows the degree of the entanglement $E(\Omega)$ as a function of the quality factor $Q_+$ and the quantum cooperativity $C_q^+(=C_+/n_{\rm th}^+)$ without the Wiener filter.
The thermal phonon number is $n_{\text{th}}^+=1$ for the left panel and $n_{\text{th}}^+=10$ for the right panel.
The red dashed line is $Q_+\epsilon=4C_+$, 
the boundary of the relation \eqref{encr}.
One can see that $Q_+\epsilon=4C_+$
well coincided with the boundary of the numerical result for the gravity-induced entanglement generation
for $C_q^+\gg1$.
The black vertical dashed line is $Q_+\epsilon=4n_{\rm th}^+$, 
the threshold of the relation \eqref{enphcr} for generating entanglement between the output lights in the Fourier space with entanglement negativity \cite{Miao20,Datta21}.
We briefly review the definition of the entanglement negativity for the output lights in Appendix B.
The gravity-induced entanglement between the output lights occurs in the region to the right of the black dashed line.

\begin{figure}[tbp]
    \centering
    \includegraphics[width=8.5cm]{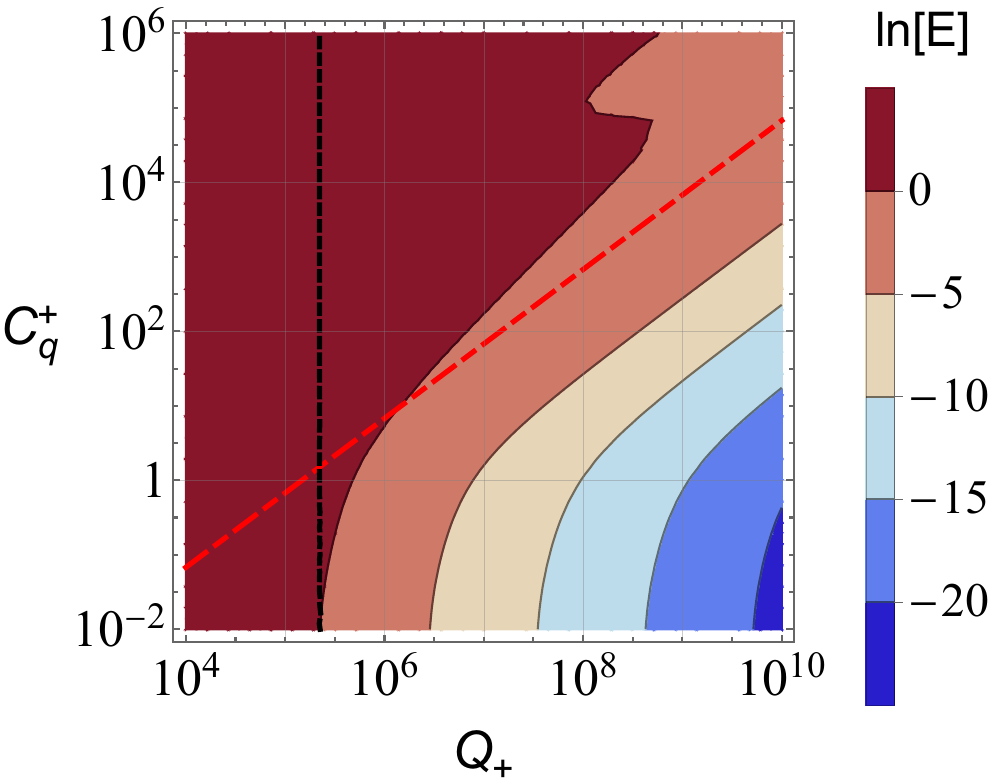}
        \hspace{0.7cm}
    \includegraphics[width=8.5cm]{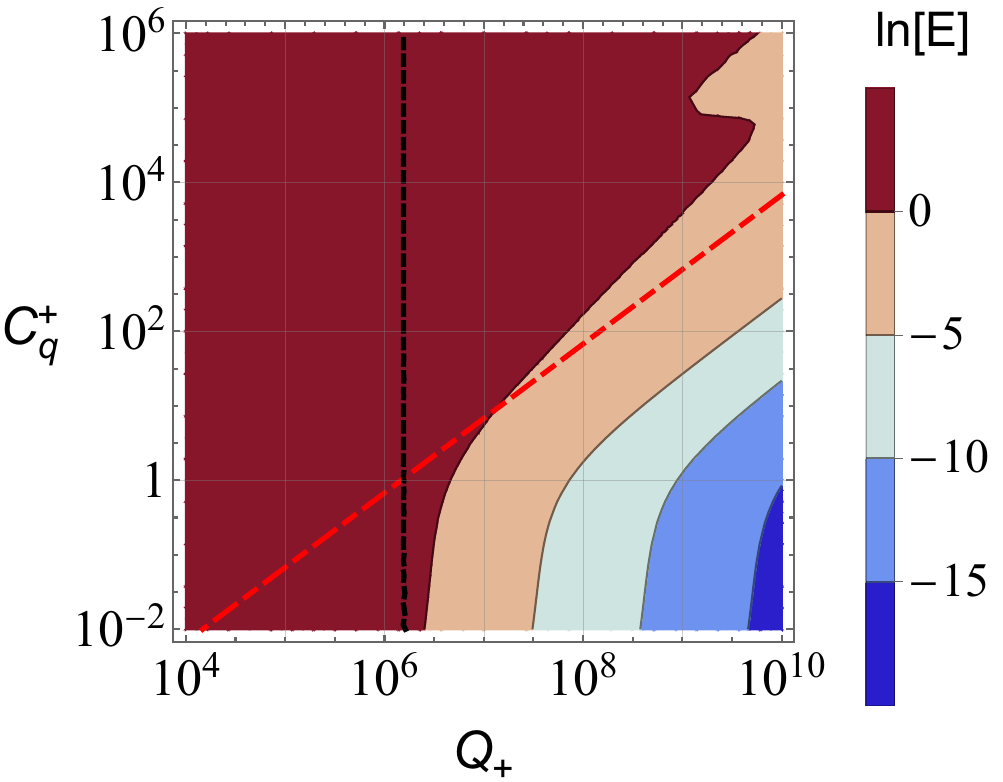}
    \caption{
    Same as Fig.~\ref{fig:ennf} but with the quantum Wiener filter.
    }
    \label{fig:en}
\end{figure}

Figure \ref{fig:en} shows the same as Fig.~\ref{fig:ennf} but with the Wiener filter. With the use of the Wiener filter,
the entanglement is more likely to occur, especially in the region of the large cooperativity and the large quality factor, 
where the degree of quantum entanglement increases by the factor $3$.
The entangled region slightly 
changes in the region of the small cooperativity and the small quality factor.
We find that the gravity-induced entanglement between the mirrors always results in the entanglement between output lights regardless of the use of the Wiener filter.

\section{
Gravity-induced squeezing}
\label{CSM}

\begin{figure}[tbp]
    \centering
    \includegraphics[width=8.7cm]{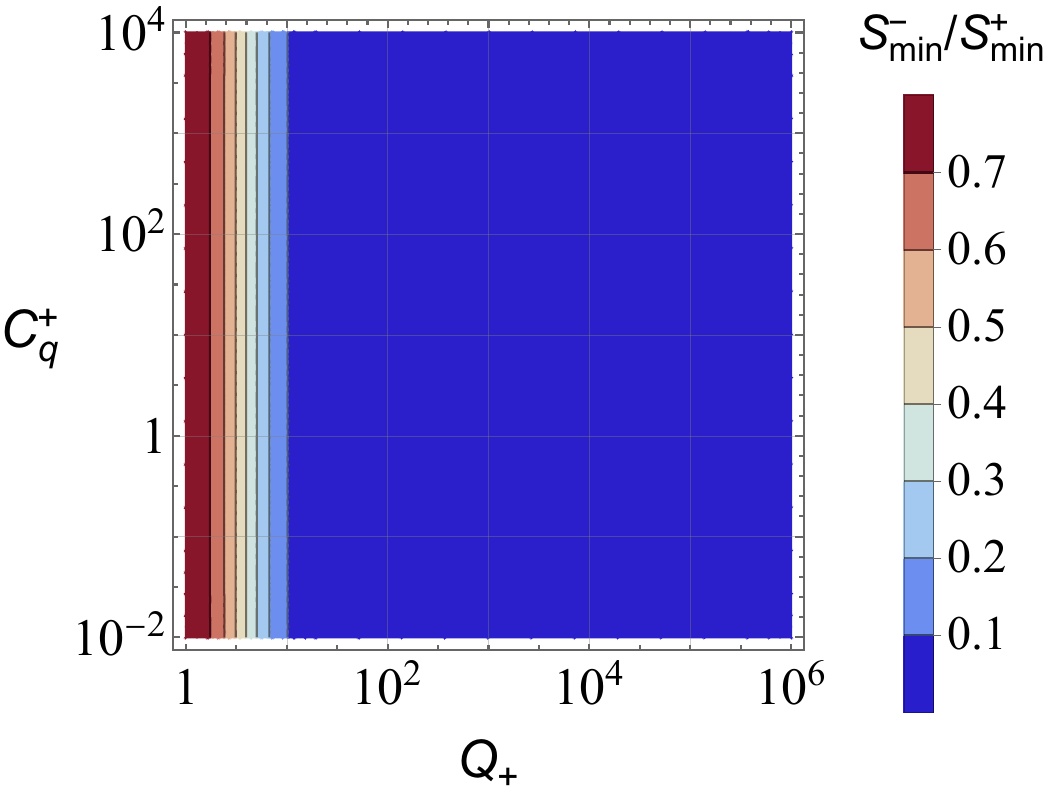}
    \hspace{0.3cm}
    \includegraphics[width=8.1cm]{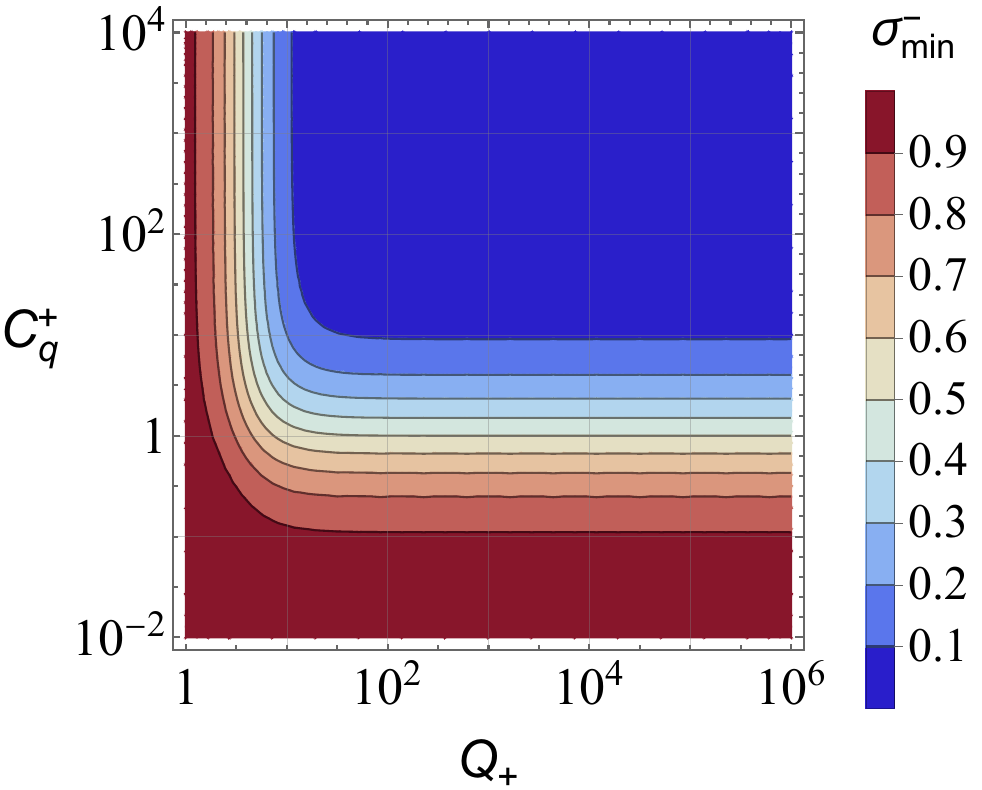}
    \caption{
    The left panel shows the ratio of the minimum variances of the mechanical differential mode to the common mode on the plane of the quantum cooperativity $C_q^+=C_+/n^+_{\rm th}$ and the quality factor $Q_+$
    without the Wiener filter.
    The right panel shows the minimum variance of the optical differential mode.
    Parameters are $\kappa/\gamma_m=10^{10}$, $\epsilon=0.27$, and $n_{\text{th}}^+=2.1\times10^{11}$.}
    \label{fig:sqme}
\end{figure}

\begin{figure}[tbp]
    \centering
    \includegraphics[width=8.7cm]{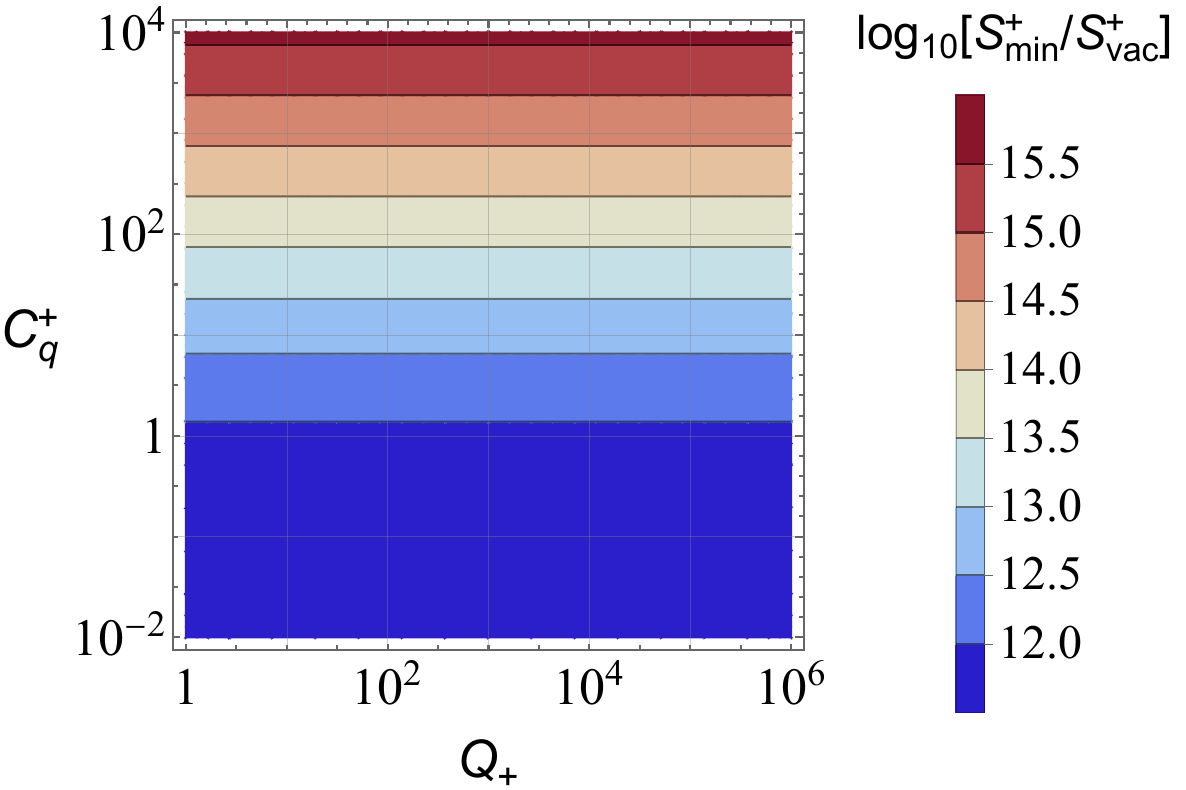}
    \hspace{0.3cm}
    \includegraphics[width=8.7cm]{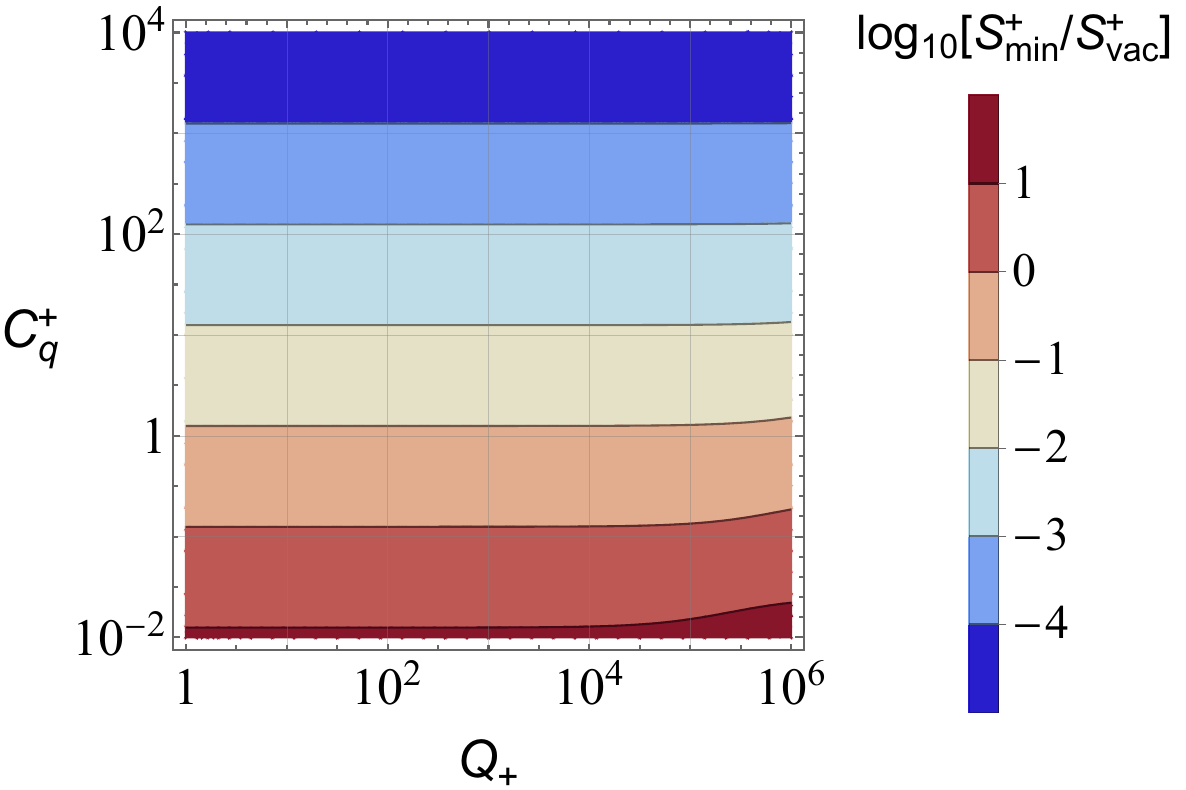}
    \caption{
    The behaviors of $S_{\rm min}^+/S_{\rm vac}^+$ without the filter (left panel) and with the filter (right panel), where the
    parameters are the same as those of Fig.~\ref{fig:sqme}.}
    \label{fig:smin}
\end{figure}

This section focuses on the gravity-induced squeezing of mirrors, which might be a quantum signature of gravity. 
Our method in this section is an application of the investigation for the gravity-induced squeezing of output lights in Ref.~\cite{Datta21} to mechanical mirrors.
We introduce the mechanical quadrature 
\begin{eqnarray}
\tilde{q}_{\theta}^\pm=\tilde{q}_\pm\cos\theta+\tilde{p}_\pm\sin\theta, 
\end{eqnarray}
where $\theta$ is a rotation angle in the phase space. The spectral density is
\begin{align}
    S_\pm(\omega,\theta)&=
    S_{\tilde{q}\tilde{q}}^\pm(\omega)\cos^2\theta
    +S_{\tilde{p}\tilde{p}}^\pm(\omega)\sin^2\theta
    +(S_{\tilde{q}\tilde{p}}^\pm(\omega)+S_{\tilde{p}\tilde{q}}^\pm(\omega))\cos\theta\sin\theta,
\end{align}
and its minimum value is
\begin{align}
    S_\pm^{\text{min}}(\omega)&=
    \frac{S_{\tilde{q}\tilde{q}}(\omega)+S_{\tilde{p}\tilde{p}}(\omega)}{2}
    -\frac{\sqrt{(S_{\tilde{q}\tilde{q}}(\omega)-S_{\tilde{p}\tilde{p}}(\omega))^2
    +(S_{\tilde{q}\tilde{p}}(\omega)+S_{\tilde{p}\tilde{q}}(\omega))^2}}{2}.
    \label{sminpm}
\end{align}

We first consider the case without the quantum filter at the mechanical resonant frequency $\omega=\Omega$.
We found the ratio of the spectral density of the differential mode to the common mode as
\begin{align}
    \label{sminme}
    \frac{S_-^{\text{min}}(\Omega)}{S_+^{\text{min}}(\Omega)}&=
    \frac{\sqrt{1-\epsilon}}{1+Q_+^2\epsilon^2},
\end{align}
which is obtained by only assuming $n_{\text{th}}^\pm=k_BT\Gamma/\hbar\Omega_\pm\gamma_m \gg1$. This result is independent of the cooperativity and the thermal noise for the mirrors, even for low quantum cooperativity, as demonstrated in the left panel of Figure~\ref{fig:sqme}, 
which plots the contour of (\ref{sminme}) on the plane of
the quality factor and the quantum cooperativity. 
In the right panel of Fig.~\ref{fig:sqme}, the
gravity-induced squeezing in the output light of the differential mode $\sigma_{\rm min}^-$ is plotted.
In Appendix \ref{gsol}, we presented a brief review of the gravity-induced squeezing in the output lights, which was first discussed in Ref.~\cite{Datta21}. 
As shown in Appendix \ref{gsol}, the gravity-induced squeezing in the output lights of the common mode is 
always unity, $\sigma_{\rm min}^+=1$. Then, the right panel of Fig.~\ref{fig:sqme} plots $\sigma_{\rm min}^-/\sigma_{\rm min}^+$, the ratio of the gravity-induced squeezing of the differential mode to the common mode. 
In Fig.~\ref{fig:sqme}, we adopted the parameters, 
 $n_{\text{th}}^+=2.1\times10^{11}$ and $\epsilon=0.27$, 
 obtained by adopting $T=1{\rm K}$, $\Gamma/2\pi=10^{-6}{\rm Hz}$, 
$\gamma_m/2\pi=10^{-4}{\rm Hz}$, 
$\Omega/2\pi=10^{-3}{\rm Hz}$,
$m/L^3=\rho\Lambda=40\times 10^3{\rm kg/m^3}$.

Because the asymmetry between the differential mode and the common mode appears only through the gravitational interaction $\epsilon$, therefore, we might understand that the deviation from unity in the ratio $S_{\rm min}^-/S_{\rm min}^+$ and $\sigma_{\rm min}^-/\sigma_{\rm min}^+$ in 
Fig.~\ref{fig:sqme} comes from the gravitational interaction.
However, it should be carefully considered what gravity-induced squeezing means. 
Figure \ref{fig:smin} plots the spectral density of the mechanical common mode $S_{\rm min}^+$ normalized by the spectral density of the ground state $S_{\rm vac}^+$ as a function of the quantum cooperativity and the quality factor,
where the spectral density of the vacuum state is obtained as 
\begin{eqnarray}
S_{\text{vac}}^+=\frac{1}{2}\left(S_{qq,\text{vac}}^+(\Omega)+S_{pp,\text{vac}}^+(\Omega)\right)=\frac{2}{\gamma_m},
\end{eqnarray}
which is obtained from Eq.~\eqref{sminpm} by ignoring the thermal phonon number and the optical noise input.
The left panel and the right panel of Fig.~\ref{fig:smin} 
compare the case without the quantum filter and the case with the quantum filter, respectively.  
The left panel of Fig.~\ref{fig:smin} shows that the spectral density of the mechanical common mode is significantly larger than that of the vacuum state without the quantum filter,
which means that the state of the mirror is dominated by thermal noises. 
The  right panel of Fig.~\ref{fig:smin} shows that the noises can be eliminated by the quantum filter, which generates the quantum squeezed state for $C_q^+\simgt0.1$.
We note that the panels of Fig.~\ref{fig:sqme} are obtained with no use of the quantum Wiener filter, and then the state of the mirrors is dominated by the thermal noises.
These facts suggest that we need to be careful to interpret the ratio $S_{\rm min}^-/S_{\rm min}^+$ in the left panel of Fig.~\ref{fig:sqme} as a quantum nature of gravity. 
There is no support that the gravity-induced squeezing in the optomechanical mode comes from the quantum nature of gravity.

\begin{table}[]
  \centering
  \begin{tabular}{c|c|cll}
   \hline
   & \multirow{3}{*}{~Oscillators~} & \multicolumn{3}{c}{~Optomechanical systems~} \\ \cline{3-5} 
   \multicolumn{1}{l|}{} & & \multicolumn{2}{c|}{~Oscillators~} & \multicolumn{1}{c}{~Output lights~~} \\ \cline{3-5} 
   \multicolumn{1}{l|}{} & & \multicolumn{1}{c|}{~With filter~~} & \multicolumn{1}{c|}{~Without filter~~} & \\
   \hline
   ~Entanglement~ & \multicolumn{1}{c|}{~Ref.~\cite{Qvarfort,Krisnanda,Datta21}~~} & \multicolumn{1}{c|}{~Fig.~\ref{fig:en}~} & \multicolumn{1}{c|}{~Eq.~\eqref{encr} and Fig.~\ref{fig:ennf}~~} & \multicolumn{1}{c}{~Ref.~\cite{Miao20,Datta21} and Eq.~\eqref{enphcr}~~} \\
   ~Squeezing~ & \multicolumn{1}{c|}{~Ref.~\cite{Datta21}~~} & \multicolumn{1}{c|}{~Fig.~\ref{fig:smin}~} & \multicolumn{1}{c|}{~Eq.~\eqref{sminme} and Fig.~\ref{fig:sqme}~~} & \multicolumn{1}{c}{~Ref.~\cite{Datta21} and Eq.~\eqref{sqphcr}~~} \\
   \hline
  \end{tabular}
  \caption{
  Summary of the conditions for gravity-induced entanglement and gravity-induced squeezing.
  }
  \label{tab:}
\end{table}
\section{SUMMARY AND CONCLUSION}
We analyzed the quantum signature of gravity in optomechanical systems under quantum control.
We found the condition for generating the gravity-induced entanglement between the mechanical mirrors with/without the quantum filter in the Fourier space.
We summarized the conditions for generating gravity-induced 
entanglement and gravity-induced squeezing in Table I.
Our new finding is the column of oscillators with filter/without filter in the optomechanical systems 
in this Table.
When the quantum filter is not used, the condition for generating the gravity-induced entanglement between the mechanical mirrors is expressed as $Q_+\epsilon>4C_+$, 
as long as the condition 
$C_q^+\gg1$ is satisfied.
When the quantum filter is used, gravity-induced entanglement between the mirrors occurs under similar conditions without the quantum filter.
However, the entanglement is more likely to occur by the use of the quantum filter, especially in the region of the large cooperativity and the large quality factor, where the degree of quantum entanglement $E(\Omega)$ increases by the factor $3$.
The condition $Q_+\epsilon>4n^+_{\rm th}$ is the one for generating the gravity-induced entanglement between the output lights \cite{Miao20,Datta21}.
As shown in Fig. \ref{fig:ennf}, the condition for generating the entanglement between the mirrors found in the present paper is more severe than the condition for the entanglement between the output lights.
This is true even with the quantum filter, as shown in Fig. \ref{fig:en}.
Hence, the gravity-induced entanglement between the output lights may be easier to detect than that between the mirrors.
A feasible combination of the parameters for generating gravity-induced entanglement is
\begin{eqnarray}
&& Q_+\simeq  10^6
\biggl(\frac{\Omega_+/2\pi}{0.1~{\rm Hz}}\biggr)\biggl(\frac{ 10^{-7}~{\rm Hz}}{\gamma_m/2\pi}\biggr),\\
&&{Q_+\epsilon}\simeq 27
\biggl(\frac{\Omega_+/2\pi}{0.1~{\rm Hz}}\biggr)^{-1}\biggl(\frac{ 10^{-7}~{\rm Hz}}{\gamma_m/2\pi}\biggr)
\biggl(\frac{\rho}{20{\rm g/cm^3}}\biggr)\biggl(\frac{ \Lambda}{2}\biggr),
\\
&&n_{\rm th}^+\simeq 6.2 
\biggl(\frac{\Omega_+/2\pi}{0.1~{\rm Hz}}\biggr)^{-1}\biggl(\frac{\Gamma/2\pi}{3\times 10^{-15}~{\rm Hz}}\biggr)
\biggl(\frac{10^{-7}~{\rm Hz}}{\gamma_m/2\pi}\biggr)
\biggl(\frac{T}{10^{-3}~{\rm K}}\biggr),
\end{eqnarray}
as long as $C_+\simlt7$, where we used $m/L^3=\rho\Lambda$ \cite{Miao20,Datta21}.
The feasibility of realizing these parameters for
generating the gravity-induced entanglement will be discussed in the future. 

We also investigated the conditions for gravity-induced squeezing of the mirrors in Sec.\ref{CSM}, 
which should be compared with the gravity-induced squeezing of optical modes \cite{Datta21}.
We found the ratio of the gravity-induced squeezing of mechanical modes, $S_-^{\text{min}}/S_+^{\text{min}}=\sqrt{1-\epsilon}/(1+Q_+^2\epsilon^2)$,
which is similar to
that of optical modes Eq.~\eqref{olsonem} in the region $C_q^+\gg1$.
The gravity-induced squeezing appears when the thermal (classical) noises are dominant. 
Further, the ratio of the gravity-induced squeezing of the differential mode to the common mode does not depend on the quantum cooperativity. 
Such a situation implies caution to interpret gravity-induced squeezing, which appears in the thermal noises dominant regime, as a quantum nature of gravity.

\acknowledgements

We thank Nobuyuki Matsumoto, Satoshi Iso, Kiwamu Izumi for valuable discussions related to the topic of the present paper.
D. M. was supported by JSPS KAKENHI (Grant No.~JP22J21267). 
K.Y. was supported by JSPS KAKENHI (Grant No.~JP22H05263. and No.~JP23H01175).

\appendix

\section{GRAVITY-INDUCED ENTANGLEMENT WITHOUT QUANTUM FILTER}
From Eqs.~\eqref{delq}-\eqref{dely}, we can derive the degree of entanglement exactly in an analytic manner as
\begin{align}
    E(\Omega)&=
    \frac{4Q_+Q_-}{(Q_+^4+Q_-^4-2Q_+^2(Q_-^2-1))^2(1+4\Omega^2/\kappa^2)^2}(Q_+^4+Q_-^4-Q_+^2(2Q_-^2-1))\notag\\
    &\times
    (1+2C_++2n_{\text{th}}^++4(1+2n_{\text{th}}^+)\Omega^2/\kappa^2)(1+2C_-+2n_{\text{th}}^-+4(1+2n_{\text{th}}^-)\Omega^2/\kappa^2),
\end{align}
where we do not use the Wiener filter.
The two mirrors are entangled if $E(\Omega)<1$.

\section{ENTANGLEMENT NEGATIVITY
BETWEEN OUTPUT PHOTONS}
The optical output quadratures in the Fourier space are
\begin{align}
    \hat{x}_\pm^{\text{out}}(\omega)&=
    \frac{\omega-i\kappa/2}{\omega+i\kappa/2}\hat{x}_\pm^{\text{in}},\\
    \hat{y}_\pm^{\text{out}}(\omega)&=
    \frac{2ig_\pm\sqrt{\kappa}}{\omega+i\kappa/2}\delta\hat{q}_\pm
    +\frac{\omega-i\kappa/2}{\omega+i\kappa/2}\hat{y}_\pm^{\text{in}},
\end{align}
where we assumed $\eta=1$.
We introduce the optical quadratures defined in Ref.~\cite{Miao20,Datta21} as
\begin{align}
    \mathcal{X}_\pm&=
    \sqrt{\frac{\Delta\omega}{2\pi}}\hat{x}_\pm^{\text{out}}(\Omega),\\
    \mathcal{Y}_\pm&=
    \sqrt{\frac{\Delta\omega}{2\pi}}\hat{y}_\pm^{\text{out}}(\Omega),
\end{align}
where $[\mathcal{X}_\pm,\mathcal{Y}_\pm^\dagger]=2i$.
We approximated the delta function as $1/\Delta\omega$, where $\Delta\omega$ is a bandwidth.
The optical covariance matrix of the common mode and the differential mode are
\begin{align}
    \label{app:cov}
    \bm{\sigma}_\pm&=
    \begin{bmatrix}
        \sigma_{\mathcal{X}\mathcal{X}}^\pm&\sigma_{\mathcal{X}\mathcal{Y}}^\pm\\
        \sigma_{\mathcal{Y}\mathcal{X}}^\pm&\sigma_{\mathcal{Y}\mathcal{Y}}^\pm
    \end{bmatrix}
    =\frac{1}{2}
    \begin{bmatrix}
        \braket{\{\mathcal{X}_\pm,\mathcal{X}_\pm^\dagger\}}&\braket{\{\mathcal{X}_\pm,\mathcal{Y}_\pm^\dagger\}}\\
        \braket{\{\mathcal{Y}_\pm,\mathcal{X}_\pm^\dagger\}}&\braket{\{\mathcal{Y}_\pm,\mathcal{Y}_\pm^\dagger\}}
    \end{bmatrix}.
\end{align}
The optical covariance matrix of the individual output photons are
\begin{align}
    \bm{\sigma}&=
    \begin{bmatrix}
        \bm{\sigma}_A&\bm{\sigma}_{AB}\\
        \bm{\sigma}_{AB}^{\text{T}}&\bm{\sigma}_B
    \end{bmatrix}
    =\frac{1}{2}
    \begin{bmatrix}
        \bm{1}&\bm{1}\\
        \bm{1}&-\bm{1}
    \end{bmatrix}
    \begin{bmatrix}
        \bm{\sigma}_+&0\\
        0&\bm{\sigma}_-
    \end{bmatrix}
    \begin{bmatrix}
        \bm{1}&\bm{1}\\
        \bm{1}&-\bm{1}
    \end{bmatrix}.
\end{align}
We introduce the entanglement negativity as
\begin{align}
    E_N&=
    -\frac{1}{2}\log_2
    \left[\frac{\Sigma-\sqrt{\Sigma^2-4\text{det}\bm{\sigma}}}{2}\right],
\end{align}
where $\Sigma=\text{det}\bm{\sigma}_A+\text{det}\bm{\sigma}_B-2\text{det}\bm{\sigma}_{AB}$.
The entanglement between the output photons occurs if and only if $E_N>0$.

\section{SQUEEZING OF OUTPUT LIGHTS DUE TO GRAVITY}
\label{gsol}
We briefly review the gravity-induced squeezing of optical mode discussed in Refs.~\cite{Miao20,Datta21}.
For simplicity, we assume $\eta=1$ and $N_{\text{th}}=0$ in the following.
The covariance matrix for the output optical modes is given as Eq.~\eqref{app:cov} in Appendix B. Considering the quadrature of the homodyne measurement $X_\theta=x^{\text{out}}\cos\theta+y^{\text{out}}\sin\theta$, the variance is
$\sigma_{\mathcal{X}_\theta\mathcal{X}_\theta}^\pm=\sigma_{\mathcal{X}\mathcal{X}}^\pm\cos^2\theta+(\sigma_{\mathcal{X}\mathcal{Y}}^\pm+\sigma_{\mathcal{Y}\mathcal{X}}^\pm)\cos\theta\sin\theta+\sigma_{\mathcal{Y}\mathcal{Y}}^\pm\sin^2\theta$, where $\theta$ is the phase of the probe beam. The minimum value of the variance is
\begin{align}
    \sigma^\pm_{\mathcal{X}_\theta\mathcal{X}_\theta}&\ge
    \sigma^\pm_{\text{min}}=
    \frac{\sigma_{\mathcal{X}\mathcal{X}}^\pm+\sigma_{\mathcal{Y}\mathcal{Y}}^\pm}{2}-\frac{\sqrt{(\sigma_{\mathcal{X}\mathcal{Y}}^\pm+\sigma_{\mathcal{Y}\mathcal{X}}^\pm)^2+(\sigma_{\mathcal{X}\mathcal{X}}^\pm-\sigma_{\mathcal{Y}\mathcal{Y}}^\pm)^2}}{2}.
\end{align}
For the common mode, the minimum value of the variance is
\begin{align}
    \sigma^+_{\text{min}}=1,
    \label{olsone}
\end{align}
then the fluctuation is always larger than the quantum fluctuation, and the quantum squeezing does not occur. However, the minimum value of the variance of the differential mode is
\begin{align}
    \label{sqphcr}
    \sigma^-_{\text{min}}&=1-\frac{4C_+(1+2n_{\text{th}}^++2C_+)\left(\sqrt{1+Q_+^2\epsilon^2/(1+2n_{\text{th}}^++2C_+)^2}-1\right)}{1+Q_+^2\epsilon^2},
\end{align}
where $\sigma_{\text{min}}^-$ is always smaller than the quantum fluctuation.
Assuming $C_+/n_{\text{th}}^+\gg1$, $C_+\gg1$, and $Q_+\epsilon\ll C_+$, the minimum value of the variance is
\begin{align}
    \label{sigmin}
    \sigma_{\text{min}}^-&=
    \frac{1}{1+Q_+^2\epsilon^2}.
\end{align}
For the other case, $C_+/n_{\text{th}}^+\ll1$, $n_{\text{th}}^+\gg1$, and $Q_+\epsilon\ll n_{\text{th}}^+$, the minimum value is
\begin{align}
    \sigma_{\text{min}}^-&=
    1-\frac{C_+Q_+^2\epsilon^2}{1+n_{\text{th}}^+Q_+^2\epsilon^2},
\end{align}
which reduces to
\begin{align}
    \sigma_{\text{min}}^-&=
    1-\frac{C_+}{n_{\text{th}}^+},
    \label{olsonem}
\end{align}
under the additional assumption $n_{\text{th}}^+Q_+^2\epsilon^2\gg1$.

\end{document}